\documentclass{PoS}

\pdfoutput=1
\usepackage[sort&compress,square, comma]{natbib}

\newcommand{\swift}{\textit{Swift}}
\newcommand{\fermi}{\textit{Fermi}}
\newcommand{\gev}{\,GeV}
\newcommand{\tev}{\,TeV}
\newcommand{\mev}{\,MeV}
\newcommand{\kev}{\,keV}

\title{The search for short-term flares in extended VHE Crab Nebula observations with the Whipple 10\,m telescope}

\ShortTitle{Crab Nebula observations with the Whipple 10\,m telescope}

\author{\speaker{A. O'Faol\'ain de Bhr\'oithe}~for the VERITAS collaboration\thanks{veritas.sao.arizona.edu}\\
        DESY, Platanenallee 6, 15738 Zeuthen, Germany\\
        E-mail: \email{anna.ofaolain.de.bhroithe@desy.de}}

\abstract{In 1989, the Whipple 10\,m telescope achieved the first indisputable detection of a TeV gamma-ray source, the Crab Nebula. Until its decommissioning in 2011, the Whipple telescope took regular measurements of the nebula. With the recent discovery of GeV gamma-ray flaring activity in the Crab Nebula, it is an opportune time to return to the Whipple telescope data set and search its extensive archive for evidence of TeV flares. A data set on the Crab Nebula spanning ten years, 2000 - 2010, is compiled and searched for day-scale flaring activity using a Bayesian-block binning algorithm. No evidence for significant flaring activity is found. Monte Carlo simulations show that low levels of flux increase on short timescales are difficult to detect. Assuming a flare duration of seven days, 99\% confidence level upper limits are calculated for the possible frequency of five-fold, two-fold and 1.5-fold flares in the data set. An upper limit of 0.02 flares per year is found for the five-fold flare, and a limit of 0.27 flares per year is placed on the two-fold flare. The detection of the 1.5-fold flare is consistent with the false-positive rate of the method, and so cannot be excluded.}

\FullConference{The 34th International Cosmic Ray Conference,\\
		30 July- 6 August, 2015\\
		The Hague, The Netherlands}

\begin{document}

\section{Introduction}
The supernova explosion that created the Crab Nebula and its central power source, a pulsar, was first observed in 1054\,AD. The Crab Nebula was the first source detected at TeV energies \cite{1989ApJ...342..379W} and its discovery laid the foundation for the field of very-high-energy (VHE; $E>100$\gev{}) astrophysics. It has been considered a standard VHE reference source, similar to its role in X-ray and high-energy gamma-ray astronomy, e.g., \cite{1974AJ...79..995T,2005SPIE...5898..22K}.

\begin{figure}[bt]
\centering
\includegraphics[width=0.5\textwidth]{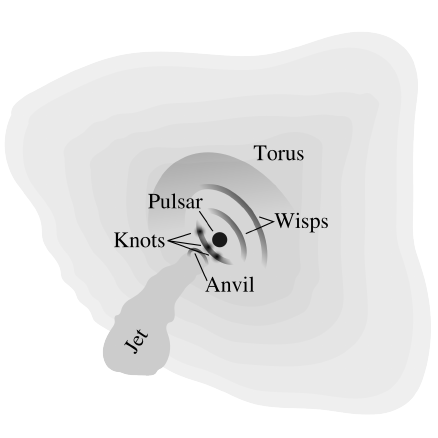}
\caption{Schematic illustration of the characteristic features of the Crab nebula seen at optical and X-ray energies.}
\label{neb}
\end{figure}

In the late 1960s, radio and optical pulsations were detected from what is now known to be the Crab pulsar \cite{1969Natur.221..525C,1968Sci...162.1481S} with a spin-down rate of 36\,ns per day \cite{1969Natur.222..551R}. A magnetohydrodynamic model was developed using the pulsar as the energy source of the nebula \cite{1974MNRAS.167....1R,1984ApJ...283..694K,1984ApJ...283..710K}. This model was capable of accounting for most of the observed features (see Figure~\ref{neb} for a schematic illustration) and predicted constant emission from the nebula.

In 2011, \cite{2011ApJL...727..40} reported a decline of $\sim7$\% in the Crab Nebula flux in the 15\,--\,50\kev{} band from 2008 to 2010 observed with the \fermi{}-GBM and independently confirmed with the \swift{}-BAT, the RXTE-PCA and the INTEGRAL-IBIS. The same publication reported a similar decline in the 3\,--\,15\kev{} band (RXTE-PCA), and in the 50\,--\,100\kev{} band (\fermi{}-GBM, \swift{}-BAT and INTEGRAL-IBIS). Contemporaneous with the publication of the X-ray variability, the discovery of short-term flares on the timescale of days with the \fermi{}-LAT \cite{2011Sci...331..739A} and AGILE \cite{2011Sci...331..736T} was announced. The \fermi{}-LAT team reported $>100$\mev{} flares detected in 2009 February and  2010 September, and the AGILE team reported flares in the range 0.1\,--\,10\gev{} in 2007 October and 2010 September, the latter being the same flare as that observed with the \fermi{}-LAT. In April of 2011, the \fermi-LAT detected the strongest gamma-ray flare seen to date \cite{2012ApJ...749...26B}. This flare lasted approximately nine days and reached a peak flux corresponding to a 30-fold increase in flux above 100\mev{}. 

The discovery of large flux increases on timescales of days and shorter was completely unexpected in the framework of the existing theoretical model. New models have been proposed to explain the newfound behavior, often suggesting regions of the nebula corresponding to the optical and X-ray features as the acceleration sites. In general, the models suggest one of three scenarios: instabilities in the termination shock, i.e., the inner knot feature e.g., \cite{2011MNRAS.414.2017K}, particle acceleration due to magnetic reconnection events in the nebula e.g., \cite{2011MNRAS.414.2229B}, or plasma kink instabilities e.g., \cite{2012Natur.482..379M} which may provide a framework for variability originating in the anvil region. Of these, only the model of \cite{2011MNRAS.414.2229B} provides predictions for the VHE domain.

Both MAGIC and VERITAS reported no change in flux or spectral shape of the VHE Crab Nebula during the \gev{} flare in 2010 September \cite{2010ATel.2967....1M,2010ATel.2968....1O}. However, ARGO-YBJ reported a flux about 3-4 times higher than the average emission at a median energy of about 1\tev{} with a statistical significance of $\sim4\,\sigma$, based on a preliminary analysis of their data \cite{2010ATel.2921...1V}. A more recent publication sets the posttrials significance of this increase to be of the order of $\sim2\,\sigma$ \cite{0004-637X-798-2-119}. This publication also reports the ARGO-YBJ long-term light curves of the Crab Nebula to be consistent with a uniform flux for median energies of 0.76, 1.8, and 5.1\tev{} respectively. Even though the rate variations in the ARGO-YBJ light curves appear to be consistent with statistical fluctuations, a correlation study of the variation in the event rate measured by ARGO-YBJ (median energy 0.76\tev{}) and the flux measured by the \fermi{}-LAT (above 100\mev{}) shows a Pearson correlation coefficient of $0.56 \pm 0.22$, which could hint at similar long-term behavior of the gamma-ray emission at $\sim$100\mev{} and $\sim$1\tev{}.

During another GeV gamma-ray flare from the Crab Nebula in 2013 March, in which a 20-fold increase in flux above $100$\mev{} for the synchrotron component of the nebula was detected \cite{2013ApJ...775L..37M}, simultaneous VHE observations were taken by VERITAS \cite{2014ApJ...781L..11A} and H.E.S.S. \cite{2014A&A...562L...4H}. Neither instrument detected evidence of any VHE flux enhancement. Since the discovery of the flares, many other simultaneous multiwavelength observations across the electromagnetic spectrum have been obtained in order to search for correlated emission at different energies, but no such correlation has been found e.g., \cite{2011A&A...527L...4B,2011ApJ...741L...5S,2013ApJ...765...52S,2012ApJ...749...26B,2015MNRAS.446..205B}.

Coverage of the GeV flares with the Whipple 10\,m telescope is unfortunately sparse, with only two days of overlap at the very end of the 2007 October flare (showing nothing unusual) and no overlap with the 2009 February flare. However, it is reasonable to assume that there have been flares from the nebula prior to the current gamma-ray satellite era. The Whipple telescope has a large archive of VHE data on the Crab Nebula dating back to the initial detection of the Crab Nebula with the telescope in 1989. In this work, a ten-year data set collected with the Whipple 10\,m telescope on the Crab Nebula from 2000 to 2010 is searched for flaring activity. Throughout this period, the configuration of the Whipple telescope remained relatively consistent from year to year, making it particularly suitable for variability studies.

The 2000\,--\,2010 Whipple telescope observations are described in \S\,\ref{sec:obs}. A Bayesian-block binning algorithm is used to search for variability in the data set, and is presented along with the results in \S\,\ref{sec:bb}. A discussion of these results follows in \S\,\ref{sec:disc}.

\section{Observations and event reconstruction} \label{sec:obs}
The Whipple 10\,m telescope \cite{2007APh...28..182K} was the first large, purpose-built telescope for atmospheric-Cherenkov astronomy. Located at the Fred Lawrence Whipple Observatory in southern Arizona at an altitude of $\sim$2.3\,km above sea level, it was completed in 1968 and operated as an imaging atmospheric-Cherenkov telescope (IACT), i.e., with an imaging camera, from 1982 to 2011, after which it was decommissioned. It was sensitive to photons with energies up to 20\tev{}, with a nominal low-energy threshold (which remained constant over the ten-year period of interest) of $\sim400$\gev{} for a Crab Nebula-like spectrum.

The telescope was of Davies-Cotton design, comprising a 10\,m diameter reflector with a radius of curvature of 7.3\,m and the camera at the focal plane. The reflector was composed of 248 tessellated hexagonal mirrors with a radius of curvature twice that of the overall dish. The camera was upgraded many times over the lifetime of the telescope. The two camera configurations used in this work each had a trigger field of view of $\sim2.6^\circ$. The different cameras and their properties are summarized in \cite{2007APh...28..182K}.

The telescope system was shut down between the end of June and the start of September each year to protect the electronic equipment from lightning damage due to the monsoon season in Arizona. As a result, an observing ``season'' is defined as beginning in September and lasting approximately until June. The Crab Nebula is visible at zenith angles $< 35^\circ$ from the Whipple Observatory site between August and April each year, allowing good quality observations of the source for up to seven months each observing season. The telescope did not operate for seven nights around the full moon period, resulting in a seven-day gap in each lunar cycle ($\sim$28 days).

A data set was compiled using Crab Nebula observations taken with the Whipple 10\,m telescope during the years 2000\,--\,2010. Data were taken in 28-minute observations in \textit{paired mode}, in which observations of the source were paired with observations of a region of the sky at the same azimuth and elevation angles with no known gamma-ray emitter for the purposes of background estimation. Data were required to be accumulated under good weather conditions and at zenith angles $< 35^\circ$, where the effective area is expected to be approximately constant. A total of $\sim$150 hours of quality-selected live time satisfied these criteria, including 328 observation pairs taken on 243 separate nights.

The data were analyzed using the standard Supercuts procedure described in Appendix B of \cite{1993ApJ...404..206R}. This analysis is a two-step process. In the first step, the images are processed to remove hardware and night-sky-background effects. This is achieved by flatfielding the images using relative-gain information from nightly calibration runs taken with a nitrogen arc lamp, screened with a diffuser to ensure uniform illumination of the camera. Software padding is used to compensate for any difference in sky brightness between the source and background regions, by injecting software noise into the events for the darker region \cite{1993tmac.conf..176C}. In the second step, the moment-fitted image parameters are calculated \cite{1985ICRC...3..445H} and the candidate gamma-ray events selected based on the principal components and derived parameters of the images.

A light curve of the Crab Nebula, binned by individual 28-minute observation and expressed in units of gamma rays per minute, is created for each observing season. The light curves are separated by season as no interseason calibration has been applied. Each light curve is fitted with a constant rate to test for temporal stability. The rate for each season is found to be consistent with a constant fit within $3\,\sigma$.

\section{Bayesian-block analysis} \label{sec:bb}

The data set is searched for variability using the Bayesian-block binning algorithm \cite{2013ApJ...764..167S}. This is a nonparametric modeling technique that is applied to the general problem of detecting and characterizing local variability in sequential data. The method is used to obtain a representation of the light curve in terms of a segmentation of the time axis into blocks --- subintervals of generally unequal size containing consecutive data that satisfies some well-defined criterion. A good example of such a criterion is the sum over all blocks of a goodness-of-fit measure of a given model of the data in each block. The optimal segmentation is that which maximizes this goodness-of-fit. In this analysis, the piecewise-constant model is used to describe the data in each block. This is the simplest model and allows for an exact treatment of the likelihood used to calculate the overall goodness-of-fit.

The form of the likelihood differs for different types of data. The type of data used in this analysis --- source rates measured at different times --- is categorized as \textit{point measurements} according to \cite[\S 1.5]{2013ApJ...764..167S}, and the corresponding form of the likelihood is used. A false-positive rate of $p_0 = 0.01$ is selected for this analysis as a compromise between a low false-positive rate and retaining the sensitivity of the test to fluctuations in the signal.

\begin{figure}[bt]
\centering
\includegraphics[scale=0.55]{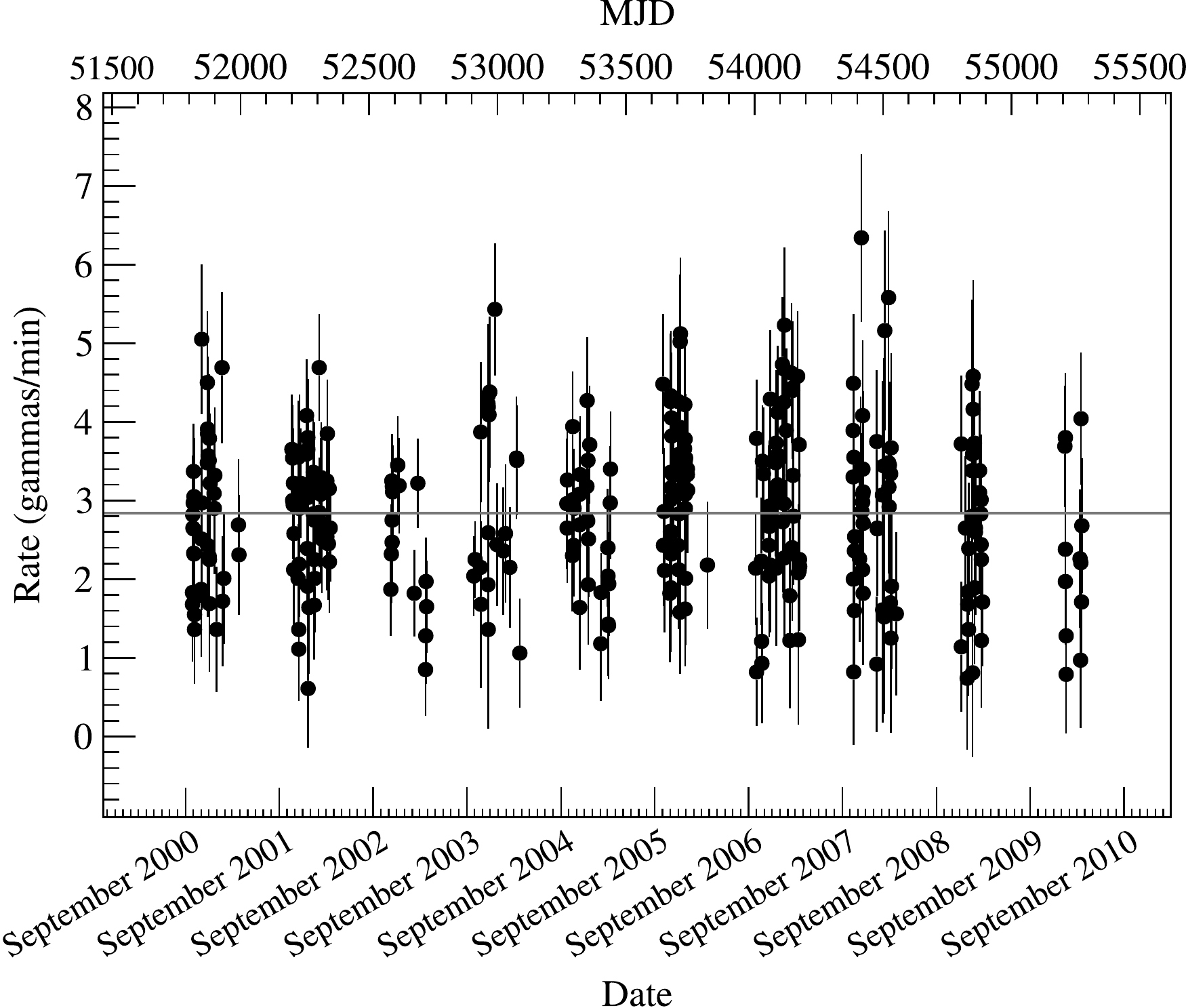}
\caption{Ten-year light curve of the Crab Nebula covering the seasons 2000\,--\,2010, binned by individual 28-minute observation. The optimal temporal binning determined by the Bayesian-block analysis is a single bin spanning the entire data set, and is indicated by the solid gray line.}
\label{whip_bb_part1}
\end{figure}

Figure~\ref{whip_bb_part1} shows the ten-season light curve of the Crab Nebula, binned by individual 28-minute observation. The result of the Bayesian-block binning on the data set is also shown. The light curve is best represented by a single time bin spanning the entire data set, which implies that the data is consistent with a constant source emission model.

\section{Discussion} \label{sec:disc}

No evidence for variability is found in ten years of Whipple telescope Crab Nebula observations. However, it is possible to obtain a limit on the level of flaring activity that could be detected. In order to do this, a toy Monte Carlo simulation was developed to simulate a single flare of known length and emission within an otherwise standard data set. Simulations of observations were created using separate Poisson distributions with constant mean for the source and background rates. The means of these distributions were set to those observed in the real data. No observations were simulated for seven consecutive nights out of each 28, mimicking the bright moonlight period. For the remaining nights, the number of Crab Nebula observations on each night was randomly chosen from an empirical distribution of the source sampling obtained by counting the number of Crab Nebula observations per night in the real Whipple telescope data (numbers range from zero to four observations of the source per night). A flare was then injected beginning on a random day, with the possibility that flares may overlap the bright moonlight period when they cannot be observed. A medium-duration flare of seven days was used for this study. It was found that on average, $\sim$30\% of the injected flares go unobserved due to a combination of the bright period and the data sampling.

To quantify the performance of the Bayesian-block algorithm, the simulation was run 10\,000 times each for flare emission levels corresponding to five times, two times, and 1.5 times the average Crab Nebula flux. The five-fold flare was detected in all simulations with at least one observation during the flare period. Using the method of \cite{2005NIMPA.551..493R}, a $99\%$ confidence level (CL) upper limit was calculated on the frequency of flares that may be present in the data set. An upper limit of $0.02$ five-fold flares per year was found, assuming at least a single observation of the flare. The two-fold flare was detected in $37\%$ of the simulations with at least one observation of the flare, leading to a $99\%$ CL upper limit of $0.27$ flares per year of this magnitude. The 1.5-fold flare was detected in $1\%$ of cases, consistent with the false-positive rate of this analysis, and so cannot be excluded. 

The only model in the literature to date with predictions for the TeV behavior of the source in the context of the flares is provided by \cite{2011MNRAS.414.2229B}. These authors suggest that the variable gamma-ray emission originates behind the shock where the pulsar wind interacts with the nebula as a result of magnetic reconnection. The model predicts TeV variability on the same timescales as those observed from the synchrotron nebula, as it is the same population of synchrotron-radiating electrons that upscatter soft photons into the VHE regime. Such variability is expected to be of the order of $\sim$10\% above 1\tev{}, and more substantial above 10\tev{}. As it is difficult to detect flares of the order of $50\%$ flux increase, a limit cannot be placed on the frequency of low level of flaring activity that may be present in the source according to this particular model.    

The nondetection of any flaring behavior in the Whipple 10\,m telescope data set is consistent with the VERITAS \cite{2014ApJ...781L..11A} and H.E.S.S. \cite{2014A&A...562L...4H} observations of the 2013 March GeV flare, which found no enhancement of the VHE flux on the level of a few percent. The detection of TeV variability in the Crab Nebula would be a very important result for determining the flare mechanism. It is possible that future instruments, especially those with a larger collection area for TeV gamma rays such as the Cherenkov Telescope Array, will be able to detect small variations in the VHE flux from the Crab Nebula on short timescales \cite{Sol2013215}, or place even stronger constraints on the level of VHE variability of the nebula during the high-energy synchrotron flares.
\vspace{2ex}

\small{
This research is supported by grants from the U.S. Department of Energy Office of Science, the U.S. National Science Foundation and the Smithsonian Institution, and by NSERC in Canada. We acknowledge the excellent work of the technical support staff at the Fred Lawrence Whipple Observatory and at the collaborating institutions in the construction and operation of the instrument. The VERITAS Collaboration is grateful to Trevor Weekes for his seminal contributions and leadership in the field of VHE gamma-ray astrophysics, which made this study possible. A.\ O'FdB acknowledges support through the Young Investigators Program of the Helmholtz Association and the support of the Irish Research Council ``Embark Initiative''.}

\bibliography{ofdb_crab}

\end{document}